\documentstyle[amsfonts,epsf,aps,twocolumn]{revtex}

\newcommand{\ee}{\end{equation}}
\newcommand{\be}{\begin{equation}}
\begin{document}
\wideabs{
\title{
Anomalous tricritical  behaviour
%scaling
 in the coil-globule 
transition of a single polymer chain}
%of a single polymer chain

\author{Annick Lesne and Jean-Marc Victor} 
  
\address{
Laboratoire de Physique Th\'eorique 
des Liquides, \\
Universit\'e Pierre et Marie Curie,  \\
Case courrier 121, 4 Place Jussieu, 75252
 Paris Cedex 05,
France\\
\vspace*{5mm}
}
\date{\today}

\maketitle
 
%\begin{abstract}
{\small
We investigate a model of self-avoiding walk exhibiting a first-order
coil-globule transition. This first-order nature, unravelled through
the coexistence of distinct coil and globule populations, has observable
consequences on the scaling properties.
A thorough analysis of the size dependence of the mean radius of gyration
evidences a breakdown of the plain tricritical scaling behaviour.
In some regimes, anomalous exponents are observed in the transition region
and  logarithmic corrections arise along 
the coexistence
curve.
\vskip 2mm\hspace{3mm}
%36.20.Ey,
 64.60.Kw, 64.90.+b
}
%\end{abstract}
%\vskip 5mm\noindent

}

%\twocolumn
Experimental observations\cite{Yosh}\cite{gittis} as well as theoretical 
studies for a specific model\cite{ILV}\cite{LV1} suggest
 that the coil-globule transition of an
isolated self-avoiding walk of finite size $N$
 might exhibit first-order features.
A strong evidence is the coexistence in some range of temperatures of
distinct coil and globule populations. Its theoretical signature is a
bimodal shape of the order parameter distribution
$P_N^{(\beta)}(t)$, where $t$ is some conformational   parameter 
of the single chain, such that the transition is achieved through an
exchange of weight between the two peaks as temperature varies.
 We here investigate the observable consequences of this first-order
feature on the scaling behaviour of the chain in the transition region.
Although one recovers a second-order transition in the infinite-size limit,
one may wonder  whether the standard tricritical picture of the
$\Theta$-point\cite{PGG}\cite{Dup} still applies.

\vskip 2mm
In  previous studies\cite{ILV}\cite{LV1}, scaling
arguments  supplemented with Monte Carlo simulations
 of a self-avoiding chain on a cubic lattice provided 
an exact analytical expression for the entropic contribution
$P_N^{(\beta=0)}(t)$ to the  distribution $P_N^{(\beta)}(t)$
of the  order parameter $t$, defined as\cite{grass}:
\be
t=\rho^{1/(\nu d-1)}=\left({N/ r^3}\right)^{5/4}
\ee 
where $r$ is the radius of gyration
of the conformation and  $\rho=N/ r^d$ its mean density
(here $d=3$ and $\nu=3/5$ is the 
Flory exponent).
We then modelled the energy of the isolated chain as $U=-NJt$.
Scaling analysis detailed in previous
 studies\cite{ILV}\cite{LV1} led to introduce rescaled variables:
\be
\hat{t}=tN^{1/n}\hspace{5mm}{\rm with}\hspace{5mm}n\approx 2
\ee
\be
\hat{\tau}=\tau N^{\phi}\hspace{3mm}{\rm with}\hspace{3mm}\phi=1-{1\over n}
\hspace{3mm}{\rm and}\hspace{3mm}
\tau=1-{\theta\over T}
\ee 
where $\theta$ is a rough estimate   of the transition temperature.
We obtained the following expression for the equilibrium distribution:
\be
\hat{P}_N(\hat{\tau},\hat{t})
=\frac{h(\hat{\tau},\hat{t})\;
e^{-A^{\prime}N^{-q(1-1/n)}\hat{t}^{-q}}}{{\cal I}_c(N,\hat{\tau})}
\ee
where
\be h(\hat{\tau},\hat{t})=\hat{t}^ce^{-A\hat{\tau}\hat{t}-B\hat{t}^n}
\ee
is the scale-invariant contribution.
The factor ${\cal I}_c(N,\hat{\tau})$
ensures that $\int_0^{\infty}\hat{P}_N(\hat{\tau},\hat{t})d\hat{t}=1$.
Let us emphasize the key role of the factor $\hat{t}^c$: for $c<-1$, this
factor, hence the function $h(\hat{\tau},\hat{t})$, are not integrable in
$\hat{t}=0$. It forbids to take the infinite-size limit in 
$\hat{P}_N(\hat{\tau},\hat{t})$ and to use the infinite-size system as 
a reference system. This factor $\hat{t}^c$ outweights the coil region and 
it will induce an observable breakdown of the scale invariance.
More generally, we introduce for any real $z$:
\be{\cal I}_{z}(N,\hat{\tau})=\int_0^{\infty}
\hat{t}^{z}e^{-A\hat{\tau}\hat{t}-B\hat{t}^n}
e^{-A^{\prime}N^{-q(1-1/n)}\hat{t}^{-q}}d\hat{t}
 \ee
An arbitrary moment simply writes:
\be\label{tt}
<\hat{t}^{\alpha}>\;={{\cal I}_{\alpha+c}(N,
 \hat{\tau})\over {\cal I}_c(N, \hat{\tau})}
\ee
Due to the factor  $\hat{t}^{z}$,
the scaling behaviour of
${\cal I}_{z}(N, \hat{\tau})$
strongly depends on the sign of $z+1$. 
For $z+1>0$, it writes:
\be
{\cal I}_{z}\propto\left\{
\begin{array}{ll}
{1\over \hat{\tau}^{z+1}}&\mbox{\rm for}\;\;\hat{\tau}\gg 1 \\
&\\
\left({-A\hat{\tau}\over nB}\right)^{z\over n-1}
e^{(n-1)B\left({-A\hat{\tau}\over nB}\right)^{n\over n-1}}
&\mbox{\rm for}\;\;\hat{\tau}<0,\;\;|\hat{\tau}|\gg 1
\end{array}
\right.
\ee
In the case where $z+1>0$, \ ${\cal I}_{z}(N, \hat{\tau})$ is thus
asymptotically scale-invariant, depending only on the rescaled variable
$\hat{\tau}=N^{\phi}\tau$ but no more on $N$.

For $z+1<0$, two regimes occur on each side of a borderline
$\hat{\tau}_{coex}(N,z)$, as shown on Figure 1:
\be\label{II}
{\cal I}_{z}\propto\left\{
\begin{array}{ll}
N^{-(z+1)(1-1/n)}&\mbox{\rm above}\\
&\\
\left({-A\hat{\tau}\over nB}\right)^{z\over n-1}
e^{(n-1)B\left({-A\hat{\tau}\over nB}\right)^{n\over n-1}}&\mbox{\rm below}
\end{array}
\right.
\ee
Here arises the above-mentionned scale-invariance breakdown. It originates
in the bimodal shape of $\hat{t}^{z-c}\hat{P}_N(\hat{\tau},\hat{t})$:
when $\hat{\tau}$ decreases, a ``coil'' peak is quite abruptly replaced
by  a ``globule'' peak after a coexistence regime precisely located around
the curve $\hat{\tau}_{coex}(N,z)$. 
Writing that the two peaks of 
$\hat{t}^{z-c}\hat{P}_N(\hat{\tau},\hat{t})$ coexist
with equal weights (which requires 
$z+1<0$) yields the following implicit equation
for the borderline $\hat{\tau}_{coex}(N,z)$:
\be 
-\phi(1+z)\ln N
=(n-1)B\left({-A\hat{\tau}\over nB}\right)^{n\over n-1}+
\ln{\cal C}_{z}+{z\ln|\hat{\tau}|\over n-1}
\ee
where ${\cal C}_{z}$ is some numerical constant.
$\hat{\tau}_{coex}(N,z)$
 could have been obtained straightforwardly by writing the balance between
the two expressions of ${\cal I}_{z}$ given in Eq.(\ref{II}).
For large $N$, $\hat{\tau}_{coex}(N,z)\propto (\ln N)^{\phi}$
for any value $z<-1$.
The borderline $\hat{\tau}_{coex}(N,z)$ thus separates
the domain in $(\hat{\tau},\ln N)$-space where the globule peak is
overwhelming, leading to a scale-invariant expression for 
${\cal I}_{z}$, and the domain where the coil peak is overwhelming.
In the latter domain, the behaviour of ${\cal I}_{z}$
is ruled by the coil peak, being controlled only by the size $N$ 
(independent of $\hat{\tau}$ at fixed $N$).
Note that we are here reasoning on the peaks of 
$\hat{t}^{z-c}\hat{P}_N(\hat{\tau},\hat{t})$ so that the two peaks cannot
be exactly identified with the coil and the globule phases,
unless $z=c$.

In the special instance where $z=c$,
 $\hat{\tau}_{coex}(N,c)$ is then the ``true'' coexistence line 
where the coil and globule populations are equally weighted;
%For $\alpha=0$, $\hat{\tau}_{coex}(N,c)$ yields the coexistence curve
%(equally weighted coil and globule populations), which
it   writes more
explicitly (denoting $\theta=\theta(\infty)$):
\be \theta(N)-\theta={\theta\;\hat{\tau}_{coex}(N,c)\over N^{\phi}}
\propto \left(\ln N\over N\right)^{\phi}\ee
Let us note that the  value $n=2$ is  consistent with 
the predicted value
$\phi=1/2$ for the crossover exponent\cite{PGG}\cite{Dup},
 since here $\phi=1-1/n$.

\vskip 2mm
Observable quantities are related to the moments of the distribution
$\hat{P}_N(\hat{\tau},\hat{t})$.
Recalling that $r=N^{1/d}t^{1/d-\nu}=N^{\nu_{\theta}}\hat{t}^{1/d-\nu}$
where: 
\be
\nu_{\theta}={1\over d}+{1\over n}\left(\nu-{1\over d}\right)
\ee
we introduce
a generalized mean radius of gyration as:
\be
R_{\alpha}=N^{1/d}<t^{\alpha}>^{-{\nu-1/d\over \alpha}}=
N^{\nu_{\theta}}<\hat{t}^{\alpha}>^{-{\nu-1/d\over \alpha}}
\ee
for any nonzero real $\alpha$.
Scaling behaviour of the moment
$<\hat{t}^{\alpha}>$ follows from the behaviour of 
${\cal I}_{c}$  and ${\cal I}_{\alpha+c}$ (see Eq.(\ref{tt})).
Of special interest are the values $\alpha=1$, as $R_1$ is related to the mean
value $<t>$, and $\alpha=-2(\nu-1/d)$, leading to the standard mean radius of
gyration $R_G=<r^2>^{1/2}$, that can be measured through
scattering experiments.
For fixed $\tau<0$, one recovers the foreseeable globule scaling law $R_{\alpha}\propto
N^{1/d}$ (here $d=3$) whereas for fixed $\tau>0$, one recovers the coil scaling
law $R_{\alpha}\propto N^{\nu}$, whatever $\alpha$ is.
In between, in the transition region $|\tau|\rightarrow 0$,
$\hat{\tau}$ finite, 
various scaling regimes are observed, depending on the 
sign of $1+c$ and $1+c+\alpha$.
 The different cases are sketched 
on Figure 2.

If $c>-1$ and $\alpha+c>-1$ (case 1), the ratio ${\cal I}_{\alpha+c}/{\cal I}_c$
is scale-invariant; it follows that $R_{\alpha}$ exhibits  the standard
tricritical behaviour:
\be
R_{\alpha}\propto N^{\nu_{\theta}} f(\hat{\tau})=
N^{\nu_{\theta}} f(N^{\phi}\tau)
\ee
with
\be
f(\hat{\tau})\propto \left\{
\begin{array}{ll}
|\hat{\tau}|^{-\left({1\over n-1}\right)
\left(\nu-{1\over d}\right)}&{\rm for}\;\;\; \hat{\tau}<0,\;|\hat{\tau}|
\gg 1\\
&\\
\hat{\tau}^{(\nu-1/d)}&{\rm for}\;\;\; \hat{\tau}>0,\;|\hat{\tau}|\gg 1
\end{array}
\right.
\ee
For  negative and large $\hat{\tau}=N^{\phi}\tau$, 
 one recovers the usual scaling behaviour 
$R_{\alpha}\propto N^{1/d}$; this crossover reflects
in the hyperscaling relation:
\be\label{hyper1}
\nu_{\theta}-\phi\,\left(\nu-{1\over d}\right)
\left({1\over n-1}\right)={1\over d}
\ee
whereas for positive and large $\hat{\tau}=N^{\phi}\tau$, 
one recovers the usual scaling behaviour 
$R_{\alpha}\propto N^{\nu}$; this crossover reflects
in the hyperscaling relation:
\be
\nu_{\theta}+\phi\,\left(\nu-{1\over d}\right)=\nu
\ee
As soon as $c\leq -1$ or  $\alpha+c\leq -1$ (cases 2 to 5),
scale-invariance breakdown and anomalous behaviour are observed.
The  scaling law for $R_{\alpha}$ still writes:
\be
R_{\alpha}\propto N^{\nu_{eff}}f(\hat{\tau})=
N^{\nu_{eff}}f(N^{\phi}\tau)
\ee
But the exponent $\nu_{eff}$ and the scaling function $f$ change when
passing
across  the curves $\hat{\tau}_{coex}(N,c)$ 
and $\hat{\tau}_{coex}(N,\alpha+c)$.
It is clear that $\phi=1-1/n$ is the crossover exponent.
In any cases (2 to 5), both  ${\cal I}_{\alpha+c}$ and ${\cal I}_c$
are scale-invariant in the leftmost region.
For $\hat{\tau}<0$ with $|\hat{\tau}|$ enough large,
any observable $R_{\alpha}$ thus exhibits the  scaling behaviour
yet encountered in case 1:
\be
R_{\alpha}\propto N^{\nu_{\theta}}\;|\hat{\tau}|^{-\left({1\over n-1}\right)
\left(\nu-{1\over d}\right)}
\ee
For $|\hat{\tau}|\gg 1$, one recovers the usual scaling behaviour 
$R_{\alpha}\propto N^{1/d}$, for any $\alpha$,
according to  the hyperscaling relation given in Eq.(\ref{hyper1}).

In cases 2 to 5,  the key point is the emergence of an intermediate region
where either ${\cal I}_{\alpha+c}$, either ${\cal I}_c$, or both
behave as $\hat{\tau}$-independent powers of $N$. This modifies the 
exponent $\nu_{\theta}$ into an anomalous exponent:
\be\label{nuprime}
\nu_{\theta}^{\prime}(\alpha)=\nu_{\theta}-
\phi\,\left({1+c\over \alpha}\right)
\left(\nu-{1\over d}\right)
\ee
observed for $\alpha>0$ or:
\be\label{nuseconde}
\nu_{\theta}^{\prime\prime}(\alpha)=\nu_{\theta}+
\phi\,\left({1+c\over \alpha}\right)
\left(\nu-{1\over d}\right)
\ee
 observed for $\alpha<0$.
The axis $\hat{\tau}=0$ plays no particular role.
In cases 2 and 3, only one curve exists, respectively
 $\hat{\tau}_{coex}(N,\alpha +c)$ and  $\hat{\tau}_{coex}(N,c)$.
On its right side (i.e. above it), $R_{\alpha}$ behaves respectively as
$N^{\nu_{\theta}^{\prime}}f_2(\hat{\tau})$
and $N^{\nu_{\theta}^{\prime\prime}}f_3(\hat{\tau})$
where $f_2$ and $f_3$ are complicated (but independent of $N$)
functions of $\hat{\tau}$.
Both reduce  to a power law for $\hat{\tau}\gg 1$:
\be
f_2(\hat{\tau})\propto 
\hat{\tau}^{-\left({c+1\over \alpha}\right)\left(\nu-{1\over d}\right)}
\hspace{10mm}
f_3(\hat{\tau})\propto \hat{\tau}^{
\left({\alpha+c+1\over \alpha}\right)\left(\nu-{1\over d}\right)}
\ee
Using definitions (\ref{nuprime}) and (\ref{nuseconde}), we obtain 
hyperscaling relations:
\be
\nu_{\theta}^{\prime}+\phi\,
\left({\alpha+c+1\over \alpha}\right)\left(\nu-{1\over d}\right)=\nu
\ee
and
\be
\nu_{\theta}^{\prime\prime}-
\phi\,\left({c+1\over \alpha}\right)\left(\nu-{1\over d}\right)=\nu
\ee
ensuring that the expected scaling behaviour 
$R_{\alpha}\propto N^{\nu}$ is recovered as soon as 
$\hat{\tau}\gg 1$.
Note that  the mean order parameter $<t>$
and the associated mean radius of gyration $R_1$  belong
to case 3 ($c=-1.13$ and $\alpha=1$).

In the last cases (4 and 5), the anomalous exponent is observed in the
intermediate but unbounded region located between
$\hat{\tau}_{coex}(N,c)$ and $\hat{\tau}_{coex}(N,\alpha+c)$.
It is still equal to  $\nu_{\theta}^{\prime}(\alpha)$ for $\alpha>0$ 
and $\nu_{\theta}^{\prime\prime}(\alpha)$ for $\alpha<0$.
In these cases, the exponent $\nu$ is observed in the rightmost region, even
if $\hat{\tau}$ is not large with respect to 1 (even negative).
Note that the mean radius of gyration $<r^2>^{1/2}$
 belongs
to case 4 ($c=-1.13$ and $\alpha+c=-1.66$).

\vskip 2mm
The remarkable  features unravelled here are:

\noindent
{\bf (i)}  \ 
the exchange of weight between the coil phase and the globule phase is far 
 sharper with respect to the variation of $\hat{\tau}$ when it occurs
through the coexistence of two peaks than when it occurs through the shift
of a simple peak.
Accordingly,  a sharp crossover is observed in the behaviour of 
${\cal I}_{z}$ if $z+1<0$, located in a narrow stripe of width 
$(\log N)^{-\phi}$ around  $\hat{\tau}_{coex}(N,z)$, whereas
a slow change with a scale-invariant 
behaviour of ${\cal I}_{z}$ is observed for $z+1>0$.

\noindent
{\bf (ii)} \  due to the $N$-dependence of the 
curves  $\hat{\tau}_{coex}(N,z)$ for $z+1<0$,
a  size-controlled crossover occurs
 at the passage across  the curves $\hat{\tau}_{coex}(N,c)$ (if $c<-1$)
or $\hat{\tau}_{coex}(N,\alpha+c)$ (if $\alpha+c<-1$).

\noindent
{\bf (iii)} \  when either $c$, either $\alpha+c$ or both
are lower than $-1$ (cases 2 to 5 of Figure 2), an intermediate region arises
where an anomalous exponent (neither $\nu$ nor
$\nu_{\theta}$) is observed.
Usual tricritical scaling is observed only for $c>-1$ and $\alpha+c>-1$
(case 1 of Figure 2).

\noindent
{\bf (iv)} \ for $c<-1$ (cases 3 to 5 of Figure 2), on the coexistence line
$\hat{\tau}_{coex}(N,c)$, the scaling
behaviour of 
$R_{\alpha}$ exhibits multiplicative logarithmic corrections
with the same exponent whatever $\alpha$ is:
\be
 R_{\alpha}=N^{\nu_{\theta}}(\ln N)^{-{1\over 2}(\nu-1/d)}=
N^{\nu_{\theta}}(\ln N)^{-{2\over 15}}
\ee

\vskip 2mm
What we here  suggest  is that the scaling behaviour around the $\Theta$-point might be
more complex than the standard tricritical behaviour\cite{PGG}\cite{Dup}.
The peculiarity of the scaling behaviour associated with the first-order
features  of the
transition is  the occurence of two ``theta regimes''.
A first  one
is observed when the globule peak is overwhelming;
it thus  involves the exponent $\nu_{\theta}$ arising from the scale
invariance of the globule phase
and the scaling function is a power law.
A second,  anomalous one involves an observable-dependent
exponent $\nu_{\theta}^{\prime}(\alpha$) (if $c<-1$ and $\alpha>0$) or
$\nu_{\theta}^{\prime\prime}(\alpha)$
(if $\alpha+c<-1$ and $\alpha<0$)  
where $\alpha$ refers to the observable $R_{\alpha}$; this 
anomalous exponent arises from a non trivial balance between the scale
invariant globule phase and the size-controlled coil phase, 
which is required to match
their incompatible scaling behaviours.
Only the detailed analysis of equilibrium distributions like
$P_N^{(\beta)}(t)$ gives reliable  predictions about the actual coil-globule
transition experienced by a chain of finite size.  

A related result is the irrelevance of standard finite-size scaling
approaches to describe the size
dependence of the thermal coil-globule transition.
Indeed, the standard finite-size analysis used
 for first-order transitions\cite{BinderL} here failed
as the two states merge in the infinite-size limit.
On the other hand, an analysis based
on the knowledge of the infinite-size second-order transition\cite{Binder}
cannot account for the first-order features here observed in finite size.
In a word, standard finite-size scaling approaches are designed to predict
 the rounding
of the transition features in finite size but cannot introduce back the
change of nature of the transition observed here.
A novel finite-size scaling analysis should thus  be designed
to interpret experimental or numerical data\cite{LV3}.

\begin{center}
{\bf Captions}
\end{center}
\vskip 3mm
\noindent{\bf Figure 1:} 
 Scaling behaviour of ${\cal I}_{z}$
for $z+1<0$ (here $z=-1.66$, corresponding to the integral ${\cal
I}_{\alpha+c}$ involved in the mean radius of gyration $<r^2>^{1/2}$).
The two scaling regimes are separated by the curve
$\hat{\tau}_{coex}(N,z)$, depending on $z$ but of 
similar shape
for any value $z<-1$.
The curve is restricted to the  domain
$\hat{\tau}\leq \hat{\tau}_g\approx -3.2$ (independent of $N$)
where  a well-identified globule state (a globule peak) exists.
For $c<-1$,  $\hat{\tau}_{coex}(N,z=c)$
is the coexistence line of the then first-order coil-globule transition.

\vskip 3mm
\noindent{\bf Figure 2:}  Critical exponents 
  of $R_{\alpha}(N,\hat{\tau})$
for the different scaling
regimes determined by the signs of $1+c$ and $1+\alpha+c$.
The bold line (cases 3,4 and 5) is the coexistence curve
$\hat{\tau}_{coex}(N,c)$ (defined for  $c+1<0$, here $c=-1.13$);
the thin line (cases 2, 4 and 5)
is the curve  $\hat{\tau}_{coex}(N,\alpha +c)$
(defined for  $1+\alpha +c<0$, here $\alpha +c=-1.66$ which 
corresponds to the mean radius of gyration $<r^2>^{1/2})$.


\begin{thebibliography}{2000}

\bibitem
[1]{Yosh} S.M. Melnikov, V.G. Sergeyev 
and K. Yoshikawa, J. Am. Chem. Soc.
{\bf 117}, p. 2401, (1995).
K.Yoshikawa and Y. Matsuzawa, Physica D, p. 220 (1995).


\bibitem
[2]{gittis} A.G. Gittis, W.E. Stites and E.A. Lattman,
{\it J. Mol. Biol.} {\bf 232}, 718 (1993).



\bibitem
[3]{ILV} J.B. Imbert, A. Lesne and J.M. Victor,
{\it Phys. Rev. E}  {\bf 56}, p. 5630  \ (1997).


\bibitem
[4]{LV1} A. Lesne and J.M. Victor, First-order $\Theta$-point of a single
polymer chain
(submitted). Preprint available at \ \ cond-mat/0004273.

\bibitem
[5]{PGG} P.G. De Gennes, 
{\it J. Phys. Lett.} {\bf 36}, L55 (1975).
{\it J. Phys. Lett.} {\bf 39}, L299 (1978).

\bibitem
[6]{Dup} B. Duplantier, {\it J. Chem. Phys.} {\bf 86}, 4233 (1987).

\bibitem
[7]{grass} Note that $t$ has also been evidenced to be relevant
order parameter of the coil-globule transition of an isolated self-avoiding
chain by
P. Grassberger and R. Hegger, {\it J. Chem. Phys.} {\bf 102}, 6881
(1995).


\bibitem
[8]{BinderL} K. Binder and D.P. Landau, 
{\it Phys. Rev. B}  {\bf 30}, 1477 (1984).

\bibitem
[9]{Binder} K. Binder, {\it Rep. Prog. Phys.}  {\bf 60}, 487 (1997).



\bibitem
[10]{LV3} A. Lesne and J.M. Victor, Novel finite-size scaling analysis
of the   coil-globule
transition
 for a single polymer chain, {\it in preparation}.

\end{thebibliography}
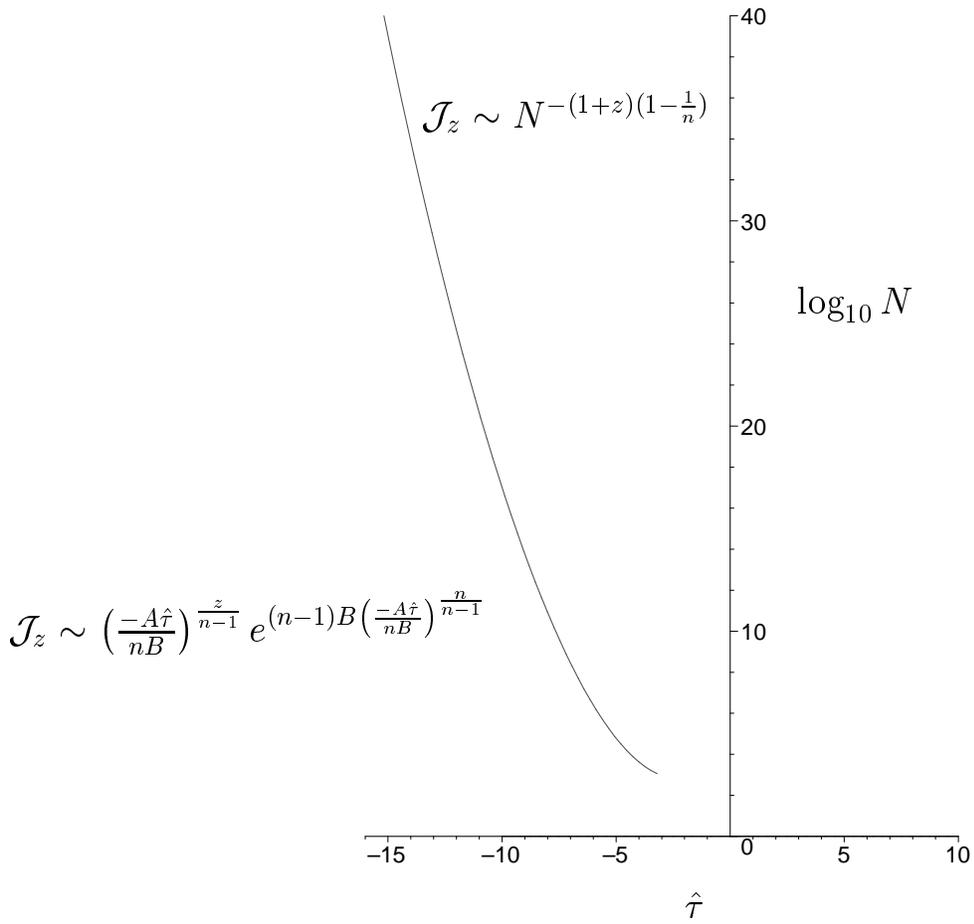
\end{document}